\documentclass{elsarticle}
\usepackage{amsmath}
\usepackage{amssymb}
\usepackage{graphicx}
\usepackage{esint}
\usepackage{bbm}
\usepackage{subfig}

\makeatletter
\def\ps@pprintTitle{%
  \let\@oddhead\@empty
  \let\@evenhead\@empty
  \let\@oddfoot\@empty
  \let\@evenfoot\@oddfoot
}
\makeatother

\begin{document}

\title{Simplicial gauge theory and quantum gauge theory simulation}

\author[tgh]{Tore Gunnar Halvorsen}
\ead{toregha@gmail.com}

\author[tms]{Torquil Macdonald S\o{}rensen}
\ead{t.m.sorensen@matnat.uio.no, torquil@gmail.com}

\address[tgh]{Department of Mathematical Sciences, Norwegian University of Science and Technology, N-7491 Trondheim, Norway}
\address[tms]{Centre of Mathematics for Applications, University of Oslo,\\ N-0316 Oslo, Norway}

\begin{abstract}
We propose a general formulation of simplicial lattice gauge theory
inspired by the finite element method. Numerical tests of convergence
towards continuum results are performed for several $SU(2)$ gauge
fields. Additionaly, we perform simplicial Monte Carlo quantum gauge
field simulations involving measurements of the action as well as
differently sized Wilson loops as functions of $\beta$.
\end{abstract}


\maketitle

\section{Introduction\label{sec:Introduction} }

\paragraph{General introduction}

Gauge quantum field theory (QFT) has been extremely successful in
modeling the behaviour of fundamental high energy particle physics.
This is done using the standard model of particle physics, which is
based on the gauge symmetry group $\mathcal{G}=U(1)\oplus SU(2)\oplus SU(3)$.
Quantum gauge field theories based on such noncommutative gauge groups
are also called Yang-Mills theories \cite{Yang:1954ek,Weinberg:1995mt,Weinberg:1996kr,Peskin:1995ev}.
Despite the massive successes of this model, there are still large
difficulties in calculating low energy properties of quarks and gluons.
When restricting to these quantum fields, the standard model reduces
to the theory of Quantum Chromodynamics (QCD), with gauge group $SU(3)$.
The problems is that through the effect of renormalization, the QCD
coupling constant increases as interaction energies is decrease, in
such a way that perturbation theory breaks down. This phenomenon is
the source of confinement in QCD. Direct paper-and-pen calculation
of masses and interactions among low energy bound states of quarks
is therefore quite problematic.

\paragraph{Lattice gauge theory}

By discretizing QCD onto a lattice, a lot of these difficulties are
removed. Lattice gauge theory (LGT) \cite{wilson74,Creutz:1984mg}
has proven itself to be a powerful method of doing nonperturbative
gauge theory calculations. It has therefore been, still is, and will
for a long time be immensely useful in testing QCD against experimental
results at low energy.

Usually LGT models are formulated using a hypercubic lattice on a
euclidean spacetime. Such a mesh preserves some discrete subgroups
of the translational, mirror and 4d rotational symmetries. Note that
a clever way of retaining continuous symmetries while working on a
lattice is to use random lattices \cite{Christ:1982ci,Christ:1982ck,Christ:1982zq}.

The models are almost always defined so as to also preserve a discrete
gauge symmetry. This has the beneficial effect of enforcing a vanishing
gluon mass in the discrete model.

\paragraph{Simplicial lattices}

Simplicial meshes have been used for QCD simulations before \cite{Drouffe:1983dk,Drouffe:1983dm,Drouffe:1983kq,Drouffe:1983vv,Cahill:1985pb,Drouffe:1984,Ardill:1983hn},
with promising numerical results. Here, we construct a simplicial
gauge theory (SGT) based on the general mathematical concept of a
simplicial complex, while preserving gauge invariance. This allows
us to define SGT on a very general class of meshes, without restricting
ourselves to a particular type of simplicial lattice.

The construction of the gauge invariant SGT action functional is inspired
by the finite element method (FEM) most commonly used for solving
partial differential equations, particularly on complicated domains
\cite{ciarlet78,monk03,nedelec1980,whitney57,hiptmair02}. The formalism
therefore includes the use of finite element function spaces on simplicial
meshes, and the concept of mass matrices. The latter has nothing to
do with physical particle masses, and is therefore not to be confused
with the usual mass matrix of quantum states within QFT.

Through the use of the FEM formulation, and the massive resources
of methods available within that subject area, we hope to gain advantages
for QCD simulations in future implementations, in particular with
regards to the possibilities of grid refinement. This could be useful
in modeling some QCD phenomena, e.g. for highly concentrated gluon
flux tubes between quarks where an increase lattice resolution might
be desired.

\paragraph{Computer simulation}

The mathematical proof of consistency between the SGT and continuous
Yang-Mills gauge theory action is described in a companion paper \cite{HalvSor2011a},
along with a description of the more comprehensive Yang-Mills-Higgs
model. In the current article we are content to provide numerical
evidence for convergence towards exact continuum results for several
choices of gauge field configurations. In addition, we perform Monte
Carlo quantum pure gauge field theory simulations for the gauge group
$SU(2)$ in temporal gauge, as a proof-of-concept for SGT. Observable
measurements include expectation values of the action density as well
as a series of different Wilson loops.

\paragraph{Outline}

Section \ref{sec:construction} contains a short repetition of the
fundamental definitions of gauge symmetry and the continuous spacetime
Yang-Mills action in subsection \ref{sec:YM_action}, the basics of
traditional lattice gauge theory in subsection \ref{sec:LGT}, as
well as an introduction to the proposed SGT action in subsection \ref{sec:SGT}.
In section \ref{sec:convergence}, we report on the numerical convergence
of the SGT action towards the exact continuum value for several different
cases of $SU(2)$ gauge fields, as well as similar results from traditional
LGT. Theoretical results proving convergence for general gauge fields
can be found in \cite{HalvSor2011a}. In section \ref{sec:MC}, we
perform Monte Carlo quantum field theory simulations in order to observe
that SGT correctly reproduces the basic aspects of the $SU(2)$ quantum
field theory. We draw our conclusions in section \ref{sec:conclusions}.
\ref{sec:simp-comp} contains a short introduction to elementary
aspects of simplicial complexes, and some notes about basis functions
and mass matrices that are use in our construction of SGT.
\ref{sec:limits} contains a calculation of strong and weak coupling
limits for a Wilson triangle and the action density. Lastly,
\ref{sec:program} contains a short discussion of some aspects of
the numerical computer implementation.

\section{Construction\label{sec:construction}}

\subsection{Continuous gauge theory\label{sec:YM_action}}

Consider the spacetime domain $\mathbb{M}=\mathbb{R}\times S$, where
$\mathbb{R}$ is time and $S\subset\mathbb{R}^{3}$. The domain $\mathbb{M}$
represents either lorentzian or euclidean spacetime, in each case
equipped with the appropriate metric. In the standard orthonormal
$\mathbb{M}$-basis $\{e_{\mu}\}_{\mu=0,1,2,3}$, a general point
$x\in\mathbb{M}$ has components $\{x^{\mu}\}_{\mu=0,1,2,3}$. Greek
indices run from $0$ to $3$, and Latin indices from $1$ to $3$.

Furthermore, in this article we shall consider pure $SU(2)$ gauge
theory. However, the construction presented is applicable to any gauge
theory based on a compact Lie group $\mathcal{G}$ which can be represented
by a subgroup of the complex unitary $n\times n$ matrices. We define
the real-valued scalar product on $\mathcal{G}$ as 
\begin{equation}
g'\cdot g:=\Re\text{tr}(g'g^{H}),\label{eq:alg_inner_prod}
\end{equation}
where $g^{H}$ is the hermitian conjugate of a matrix $g$.

The connection between the continuous theory and the discrete simplicial
theory is most easily seen in a coordinate free formulation. Thus,
we start with a coordinate free formulation, before we give the more
familiar coordinate based one.

The free variable in pure Yang-Mills theory with gauge Lie group $\mathcal{G}$
is a gauge potential or more formally a one form $A$ on $\mathbb{M}$,
with values in the corresponding gauge Lie algebra $\mathfrak{g}$.
For simplicity of notation, we hereby specify $\mathcal{G}=SU(2)$
and $\mathfrak{g}=\mathfrak{su}(2)$. We split $A$ into temporal
and spatial components $A=(A_{0},\mathbf{A})$. In this context, $A_{0}$
can be thought of as a scalar function%
\footnote{However, not a scalar in the sense of spacetime symmetry transformation
properties.%
} , and $\mathbf{A}$ as a spatial vector. The curvature (field strength)
of such a one form is given by 
\begin{equation}
F(A)=\mathbbm dA+\frac{i}{2}[A,A]=d_{0}\mathbf{A}+dA_{0}+d\mathbf{A}+i[\mathbf{A},A_{0}]+\frac{i}{2}[\mathbf{A},\mathbf{A}],\label{eq:field_strength}
\end{equation}
where $\mathbbm d=(d_{0},d)$, $d_{0}$ and $d$ denote exterior derivative
in the temporal and spatial directions respectively, and $[\cdot,\cdot]$
is the commutator between Lie algebra valued one forms. We choose
the basis $\{t^{a}\}_{a=1,2,3}$, where $t^{a}:=\sigma^{a}/2$, for
$\mathfrak{su}(2)$, where $\{\sigma^{a}\}_{a=1,2,3}$ are the Pauli
matrices. Thus, we can expand the gauge field into components, $A=A^{a}t^{a}$.
We also have 
\begin{equation}
[A,A]=\sum_{ab}A^{a}\wedge A^{b}[t^{a},t^{b}]=\sum_{abc}i\varepsilon^{abc}A^{a}\wedge A^{b}t^{c},\label{eq:commutator}
\end{equation}
where $\varepsilon^{abc}$ is the antisymmetric Levi-Civita symbol
with $\epsilon^{123}=1$ and $\wedge$ is the wedge product (exterior
product). For later convenience we split the curvature in a temporal
and spatial part 
\begin{equation}
F^{t}(A)=d_{0}\mathbf{A}+dA_{0}+i[\mathbf{A},A_{0}],\qquad F^{s}(A)=d\mathbf{A}+\frac{i}{2}[\mathbf{A},\mathbf{A}].\label{eq:field_strength_splitting}
\end{equation}

The action that defines the gauge theory is the functional 
\begin{equation}
S[A]=\frac{1}{4e^{2}}\int_{\mathbb{M}}|F(A)|^{2}=\frac{1}{4e^{2}}\int_{\mathbb{M}}|F^{t}(A)|^{2}+|F^{s}(A)|^{2},\label{cont:action}
\end{equation}
 where the norms are generated the metric and $e$ is the dimensionless
Yang-Mills coupling constant.

A gauge transformation is defined by a choice of $G(x)\in SU(2)$
for each $x\in\mathbb{M}$, and transforms the gauge field as 
\begin{equation}
A_{0}\mapsto G\left(A_{0}+d_{t}\right)G^{-1},\qquad\mathbf{A}\mapsto G\left(\mathbf{A}+d\right)G^{-1}.\label{cont:gauge_field_trans}
\end{equation}
Note that the action $S[A]$ is invariant under such gauge transformations.
For a more precise mathematical exposition, see \cite{HalvSor2011a}.

A formulation more familiar within physics is obtained by expressing
the one form and curvature in coordinates. In other words, one decomposes
the one-form $A^{a}$ in the basis $\{dx^{\mu}\}$, i.e. $A^{a}=\sum_{\mu}A_{\mu}^{a}dx^{\mu}$.
The exterior derivative of such a one-form is given by 
\begin{equation}
\mathbbm dA^{a}=\sum_{\mu\nu}\partial_{\nu}A_{\mu}^{a}dx^{\nu}\wedge dx^{\mu}=\sum_{\mu\nu}\frac{1}{2}(\partial_{\mu}A_{\nu}^{a}-\partial_{\nu}A_{\mu}^{a})dx^{\mu}\wedge dx^{\nu}.\label{eq:ext_deriv}
\end{equation}
Furthermore, the curvature is given by $F^{a}=\sum_{\mu\nu}\frac{1}{2}F_{\mu\nu}^{a}dx^{\mu}\wedge dx^{\nu}$,
where 
\begin{equation}
F_{\mu\nu}^{a}=\partial_{\mu}A_{\nu}^{a}-\partial_{\nu}A_{\mu}^{a}-\varepsilon^{abc}A_{\mu}^{b}A_{\nu}^{c}.\label{eq:field_strength_components}
\end{equation}
Finally, the action can be expressed as 
\begin{equation}
S=\frac{1}{4e^{2}}\int_{\mathbb{M}}\sum_{\mu\nu a}F_{\mu\nu}^{a}F^{a\mu\nu}dx,\label{eq:action_ym}
\end{equation}
the usual coordinate dependent expression for the Yang-Mills action
functional.

\subsection{Lattice gauge theory\label{sec:LGT}}

To see the connection between lattice gauge theory (LGT) and the simplicial
gauge theory (SGT), we will in this section give a brief overview
of the discretization procedure from LGT. For a more complete description
see e.g. \cite{Creutz:1984mg}.

The discretization procedure of both LGT and SGT is based on the following
identity. Consider a small surface $\Sigma$ with area proportional
to $h^{2}$, where $h$ is a small positive quantity. Then the following
identity holds 
\[
\oiint_{\Sigma}F(A)=\mathcal{H}(A)-1+\mathcal{O}(h^{3}),
\]
where $\mathcal{H}(A)$ is the holonomy of the one-form A, i.e. the
parallel transport induced by $A$ around the boundary of $\Sigma$.
This parallel transport is defined as follows. Given a curve $\gamma:[0,1]\rightarrow\mathbb{M}$,
such that $\gamma(0)=x$ and $\gamma(1)=y$, the parallel transport
operator along $\gamma$ is given by 
\[
U_{\gamma}(x,y)=P\left(\exp(i\int_{\gamma}A)\right),
\]
where $P$ denotes path-ordering, and the subscript $\gamma$ is attached
to $U$ to denote the path dependence. In LGT, this quantity is known
as the Wilson line.

In LGT, spacetime $\mathbb{M}$ is usually discretized by a uniform
hypercubic lattice $\mathbb{L}$. Neighbouring node positions are
related through translation vectors $\{a_{\mu}\}$ for which we assume
$|a_{\mu}|=h$ for all $\mu$. To each edge $e$ which connects neighbouring
nodes, $n$ and $n+a_{\mu}$ for some $\mu$, we attach an approximation
of the parallel transport operator along $e$. Thus, 
\begin{equation}
U_{\mu}(n):=\exp(ihA_{\mu}(n+\frac{1}{2}a_{\mu}))\approx U_{e}(n,n+a_{\mu})=P\left(\exp(i\int_{n}^{n+a_{\mu}}A)\right).\label{eq:parallel-trans-op}
\end{equation}
In LGT this quantity is called a link variable, link matrix or link
group element. Furthermore, given a face $f$ of a cube in the mesh,
called a plaquette, we approximate the holonomy associated to this
face as the path-ordered product of the link variables along its boundary.
In other words, if $f$ lies in the $\mu\nu$ plane, with nodes $n$,
$n+a_{\mu}$, $n+a_{\nu}$, and $n+a_{\mu}+a_{\nu}$, we approximate
the holonomy as 
\begin{equation}
\begin{split}
U_{f}(n):=U_{\mu\nu}(n)&:=U_{\mu}(n)U_{\nu}(n+a_{\mu})U_{\mu}^{H}(n+a_{\nu})U_{\nu}^{H}(n)\\
&\approx\mathcal{H}(A):=P\left(\exp(i\oint_{\partial f}A)\right),\label{holonomy}
\end{split}
\end{equation}
where $\partial f$ denotes the boundary of the plaquette $f$. In
LGT, this quantity is known as the Wilson loop. Moreover, we approximate
the curvature as 
\begin{equation}
F_{\mu\nu}^{f}\approx U_{f}-\mathbbm1.\label{curvature_appr}
\end{equation}
Finally, the LGT action is defined as 
\begin{equation}
S_{LGT}=\beta\sum_{f}\frac{1}{4}\text{tr}\left[(U_{f}-\mathbbm1)(U_{f}-\mathbbm1)^{H}\right]=\beta\sum_{f}1-\frac{1}{4}\text{tr}(U_{f}+U_{f}^{H}),\label{lgt_action}
\end{equation}
where $\beta$ is related to the coupling constant by $\beta=4/e^{2}$.
A discrete gauge transformation is associated with a choice of $G(n)\in SU(2)$
for each node n. Each link variable then transforms as 
\begin{equation}
U_{\mu}(n)\mapsto G(n)U_{\mu}(n)G(n+a_{\mu})^{-1}.\label{eq:discrete-gauge-trans}
\end{equation}
By the cyclic invariance of the trace, the action $S_{LGT}$ is discretely
gauge invariant.

\paragraph{Remarks}

The LGT action can be viewed as a mass lumped FEM action, and this
observation is useful to have in mind when we construct the simplicial
analogue. In the FEM setting, the gauge potential is assumed to be
a lowest order curl-conforming Nédélec element in 4d on hypercubes,
with one dimension representing time \cite{nedelec1980}. The degree
of freedom associated to such a gauge potential at an edge $e$ from
$n$ to $n+a_{\mu}$ is 
\[
A_{e}=\int_{n}^{n+a_{\mu}}A=hA_{\mu}(n+\frac{1}{2}a_{\mu}).
\]
The parallel transport operator is as in equation \eqref{eq:parallel-trans-op},
i.e. $U_{\mu}(n)=\exp(iA_{e})$. Then, the holonomy is approximated
as in equation \eqref{holonomy}, the curvature as in equation \eqref{curvature_appr},
and one considers $U_{f}-\mathbbm1$ as the components of the two-form
\[
\sum_{f}(U_{f}-\mathbbm1)\omega_{f},
\]
where $\{\omega_{f}\}$ are the Nédélec basis two-forms. The FEM action
associated to such a two form is 
\[
S:=\frac{\beta}{2}\sum_{f,f'}M_{ff'}\text{tr}\left[(U_{f}-\mathbbm1)(U_{f'}-\mathbbm1)^{H}\right],\qquad M_{ff'}:=\int_{\mathbb{M}}\omega_{f}\cdot\omega_{f'},
\]
where $M_{ff'}$ is called the mass matrix, and $(\cdot)$ denotes
the scalar product of alternating forms w.r.t. the metric. The mass
matrix is not diagonal, which means that the discrete curvature at
different faces interact. This again implies that the action is not
discretely gauge invariant. However, by diagonalizing the mass matrix
using numerical quadrature, this action reduces to the LGT action,
equation \eqref{lgt_action}. The diagonalization procedure can also
be shown to be numerical consistent in the sense of approximation
theory \cite{HalvSor2011a}.

\subsection{Simplicial gauge theory\label{sec:SGT}}

In this section we construct the discretely gauge invariant simplicial
gauge theory (SGT) action on a simplicial complex, as defined in appendix
\ref{sec:simp-comp}. The construction is the simplicial analogue
of the FEM action described above, including additional parallel transport
operators to make it discretely gauge invariant.

The curvature associated to the temporal and spatial faces is defined
exactly as in LGT. In the notation of appendix \ref{sec:simp-comp},
consider a temporal and spatial face 
\begin{equation}
\begin{split}f_{t}(\tau) & :=\{i_{\tau},j_{\tau},j_{\tau+\Delta t},i_{\tau+\Delta t}\}\\
f(\tau) & :=\{i_{\tau},j_{\tau},k_{\tau}\},
\end{split}
\label{eq:face_nodes}
\end{equation}
where $i_{\tau}$ denotes node $i$ at time $\tau$. The time dependency
will from here on often be suppressed, unless confusion can arise.
The spatial and temporal holonomies associated to these faces, induced
by the gauge potential, are approximated as 
\begin{equation}
\begin{split}U_{f_{t}}(i_{\tau}) & =U(i_{\tau},j_{\tau})U(j_{\tau},j_{\tau+\Delta t})U(j_{\tau+\Delta t},i_{\tau+\Delta t})U(i_{\tau+\Delta t},i_{\tau})\\
U_{f}(i) & =U(i,j)U(j,k)U(k,i),
\end{split}
\label{eq:face-holonomies}
\end{equation}
where the arguments $i_{\tau}$ and $i$ are included to indicate
where the holonomy is located, and the parallel transport operators
are defined exactly as in LGT, i.e. equation \eqref{eq:parallel-trans-op}.
We observe that the holonomies located at different nodes are related
through the formulas 
\[
\begin{split}U_{f_{t}}(i_{\tau}+\Delta t) & =U(i_{\tau+\Delta t},i_{\tau})U_{f_{t}}(i_{\tau})U(i_{\tau},i_{\tau+\Delta t})\\
U_{f}(j) & =U(j,i)U_{f}(i)U(i,j),
\end{split}
\]
which give formulas for parallel transport of curvature. Hence, we
have defined the curvature associated to the temporal and spatial
faces in our 4d mesh. The distinguished point of $f$ and $f_{t}$,
i.e. the location of their holonomy, are denoted $\dot{f}$ and $\dot{f}_{t}$
respectively. Note that under a discrete gauge transformation, the
parallel transport operators are transformed as in LGT, i.e. 
\[
\begin{split}U(i_{\tau},i_{\tau+\Delta t}) & \mapsto G(i_{\tau})U(i_{\tau},i_{\tau+\Delta t})G(i_{\tau+\Delta t})^{-1}\\
U(i,j) & \mapsto G(i)U(i,j)G(j)^{-1},
\end{split}
\]
for $G(i)\in SU(2)$ for each vertex $i$.

As in LGT the curvature is approximated as 
\begin{equation}
\begin{split}F^{t} & \approx U_{f_{t}}-\mathbbm1\\
F^{s} & \approx U_{f}-\mathbbm1,
\end{split}
\label{eq:SGT-curv}
\end{equation}
considered as components of the two-forms 
\[
\begin{split} & \sum_{f_{t}}(U_{f_{t}}-\mathbbm1)\Lambda_{f_{t}}\\
 & \sum_{f}(U_{f}-\mathbbm1)\Lambda_{f},
\end{split}
\]
where the $\Lambda$ are basis functions as described in appendix
\ref{sec:simp-comp}. The associated FEM action is $S=S_{t}+S_{s}$,
where the temporal part is 
\begin{equation}
S_{t}=\frac{\beta}{2}\Re\sum_{f_{t},f_{t}'}M_{f_{t}f_{t}'}\text{tr}\left[(U_{f_{t}}-\mathbbm1)(U_{f_{t}'}-\mathbbm1)^{H}\right],\qquad M_{f_{t}f'_{t}}:=\int_{\mathbb{M}}\Lambda_{f_{t}}\cdot\Lambda_{f'_{t}},\label{eq:sgt-action-int-temp}
\end{equation}
and the spatial part is 
\begin{equation}
S_{s}=\frac{\beta}{2}\Re\sum_{f,f'}M_{ff'}\text{tr}\left[(U_{f}-\mathbbm1)(U_{f'}-\mathbbm1)^{H}\right],\qquad M_{ff'}:=\int_{\mathbb{M}}\Lambda_{f}\cdot\Lambda_{f'},\label{eq:sgt-action-int-spat}
\end{equation}
where $\beta=2/e^{2}$. Note that we have suppressed the dependency
of $S$ on $A$. Again, $M_{f_{t}f'_{t}}$ and $M_{ff'}$ are called
mass matrices that depend on the details of the mesh, and are described
more in detail in appendix \ref{sec:simp-comp}. As pointed out in
the discussion about the FEM formulation of LGT, the mass matrices
are not diagonal. This implies that the action is not discretely gauge
invariant. However, this can be resolved by parallel transport of
curvature. The temporal and spatial part of the action, $S_{t}$ and
$S_{s}$, are now treated separately.

\paragraph{The temporal part}

Let $f_{t}(\tau)$ and $f'_{t}(\tau)$ be two temporal faces. We now
use some properties of the basis functions, which are explained in
appendix \ref{sec:simp-comp}. Since the temporal basis face functions
$(\Lambda_{f_{t}})$ are piecewise constant in time, the interactions
between the temporal curvature occur only at coinciding time intervals.
Also, by properties of the edge basis functions $(\lambda_{e})$,
which define the temporal basis face functions, we can connect the
curvature at $f_{t}$ with the curvature at $f_{t}'$ by parallel
transport along at most one edge. Thus, we connect the curvatures
by parallel transport along the connecting edge $e=\{\dot{f}_{t},\dot{f}_{t}'\}$
of their distinguished points. In other words, we approximate the
temporal part of the action by 
\begin{equation}
\begin{split}
S_{SGT}^{t}:=\frac{\beta}{2}\Re&\sum_{f_{t}(\tau),f'_{t}(\tau)}M_{f_{t}(\tau),f_{t}'(\tau)}\times\\
&\times\text{tr}\left[U(\dot{f}_{t}',\dot{f}_{t})\left(U_{f_{t}(\tau)}-\mathbbm1\right)U(\dot{f}_{t},\dot{f}_{t}')\left(U_{f_{t}'(\tau)}-\mathbbm1\right)^{H}\right].\label{eq:SGT-action-temp}
\end{split}
\end{equation}

\paragraph{The spatial part}

Let $f$ and $f'$ be two spatial faces of a tetrahedron $T$. The
curvature associated to the face $f$ at time $\tau$ will interact
with the curvature associated to the face $f'$ not only at time $\tau$,
but also at times $\tau\pm\Delta t$, since the facial basis functions
are piecewise affine in time. Thus, to connect the curvature at $f(\tau)$
with the curvature at $f'(\tau')$ we must parallel transport in both
space and time. Thus, we replace 
\[
(U_{f(\tau)}-\mathbbm1)(U_{f'(\tau')}-\mathbbm1)^{H}
\]
by 
\[
U(\dot{f}'(\tau),\dot{f}(\tau))(U_{f(\tau)}-\mathbbm1)U(\dot{f}(\tau),\dot{f}'(\tau))U(\dot{f}'(\tau),\dot{f}'(\tau'))(U_{f'(\tau')}-\mathbbm1)^{H}U(\dot{f}'(\tau'),\dot{f}'(\tau))
\]
in the FEM action \eqref{eq:sgt-action-int-spat}. In words, we first
parallel transport the curvature associated to $f$ , located at the
vertex $\dot{f}(\tau)$) to the vertex $\dot{f}'(\tau)$ along the
edge $e=\{\dot{f}(\tau),\dot{f}'(\tau)\}$. Then we parallel transport
it in the temporal direction from $\dot{f}'(\tau)$ to $\dot{f}'(\tau')$.
So, we approximate the spatial part of the action as 
\begin{equation}
\begin{split}S_{SGT}^{s}:=\frac{\beta}{2}\Re\sum_{f(\tau),f'(\tau')} & M_{f(\tau),f'(\tau')}\text{tr}\left[U(\dot{f}'(\tau),\dot{f}(\tau))\left(U_{f(\tau)}-\mathbbm1\right)U(\dot{f}(\tau),\dot{f}'(\tau))\right.\\
 & \left.\times U(\dot{f}'(\tau),\dot{f}'(\tau'))\left(U_{f'(\tau')}-\mathbbm1\right)^{H}U(\dot{f}'(\tau'),\dot{f}'(\tau))\right].
\end{split}
\label{eq:SGT-action-spat}
\end{equation}
The simplicial gauge theory action is then defined as 
\begin{equation}
S_{SGT}:=S_{SGT}^{t}+S_{SGT}^{s},\label{eq:SGT-action}
\end{equation}
and by the cyclic invariance of the trace, this action is discretely
gauge invariant. A companion paper \cite{HalvSor2011a} contains
more details about this construction, as well as mathematical proofs
of consistency with the continuous action \eqref{cont:action} in
the sense of approximation theory.

\section{Computer simulation\label{sec:comp-sim}}

For our SGT computer simulations, we chose the euclidean cubic domain
$\mathbb{M}=[0,1]^{4}\subset\mathbb{R}^{4}$ with periodic boundary
conditions. We simulated the pure gauge SGT action \ref{eq:SGT-action}
in temporal gauge on a simplicial lattice with the gauge group $SU(2)$.
Choice of gauge is not necessary, but it does simplify the algorithm
slightly, since all temporal edge matrices then reduce to the identity.

\begin{figure}
\begin{centering}
\includegraphics[width=6cm]{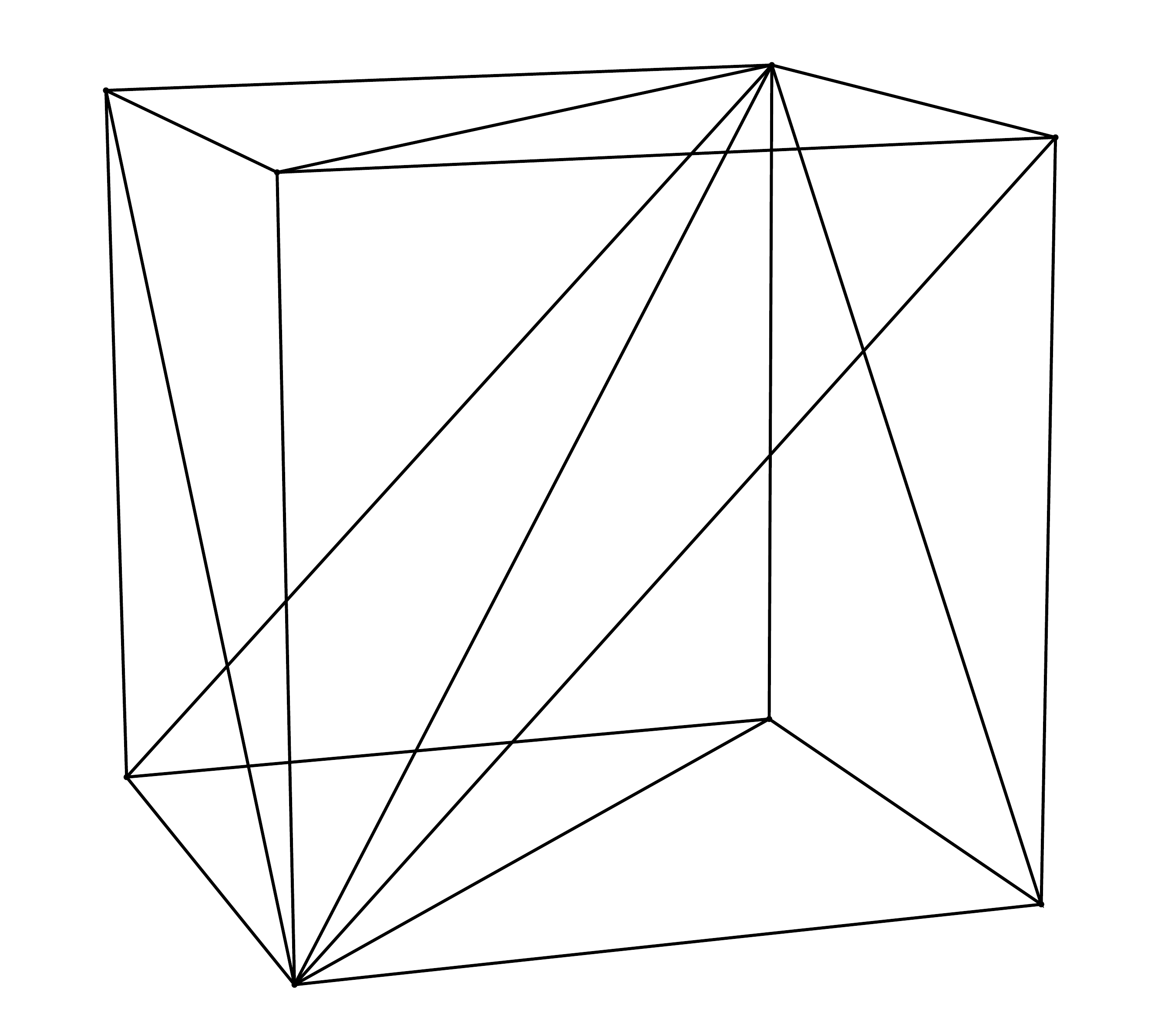}
\par\end{centering}

\caption{Elementary 3d mesh building block containing six tetrahedra, all of
which share the single interior diagonal.\label{fig:building-block}}
\end{figure}

The spatial lattice was constructed using a cubic arrangement of $N^{3}$
identical building block cubes of size $h^{3}$, each consisting of
six tetrahedra as shown in figure \ref{fig:building-block}. The resulting
spatial mesh was repeated at $N$ consecutive time steps to form a
cubic domain of physical volume $(hN)^{4}$. As described above, each
spatial edge is part of two temporal square-shaped faces, going forward
and backward in time.

The SGT action employs parallel transport matrices in order for gauge
invariance to be respected. By defining the distinguished points of
all spatial and temporal faces to coincide for as many pairs of faces
as possible, we only need the parallel transport matrices for terms
in the action involving pairs of temporal faces with no common nodes.
More details regarding the exact computer implementation are given
in appendix \ref{sec:program}.

\subsection{Convergence of the action\label{sec:convergence}}

In order to check the continuum limit of the discrete action, we examined
four different gauge field configurations for which the exact continuum
value $S_{\mbox{cont}}$ of the action is calculable. We did numerical
calculations for square meshes with $N=4,8,16,32$ in order to observe
convergence of the numerical values towards the exact values. By the
estimates in \cite{HalvSor2011a} we expect that the error be of
second order in the lattice constant $h$. We used the following gauge
field configuration cases:
\begin{enumerate}
\item Gauge field oriented towards the $x$-direction in space and towards
$t^{3}$ within $\mathfrak{su}(2)$, with a sinusoidal $t$-dependence.
The only nonzero component of the gauge field $A$ is 
\[
A_{x}^{3}(t,x,y,z):=\frac{e}{2\pi}\sin(2\pi t),\quad S=1.
\]
 
\item Gauge field oriented towards the $y$-direction in space and $t^{3}$
within $\mathfrak{su}(2)$, with a sinusoidal $x$-dependence. The
nonzero component of the gauge field in this case was 
\[
A_{y}^{3}(t,x,y,z):=\frac{e}{2\pi}\sin(2\pi x),\quad S=1.
\]
 
\item A case with two nonzero components, 
\[
A_{x}^{1}:=\frac{e}{2\pi}\sin(2\pi y),\quad A_{y}^{2}:=\frac{e}{2\pi}\sin(2\pi x),\quad S=\frac{1}{2}+\frac{e^{2}}{8(2\pi)^{4}}.
\]

\item A constant field that only contributes to the nonlinear term in the
field strength,
\[
A_{x}^{1}:=\sqrt{e},\quad A_{y}^{2}:=\sqrt{e},\quad S=\frac{1}{2}.
\]

\end{enumerate}
In order to provoke a sizable nonlinear contribution case 3, we chose
a small $\beta=2/e^{2}=1/5$. The link matrices needed to evaluate
the SGT action are calculated from these gauge fields by means of
the exponential map \eqref{eq:parallel-trans-op}.

\begin{figure}
\subfloat[LGT\label{fig:LQCD-action-error}]{\includegraphics[width=6cm]{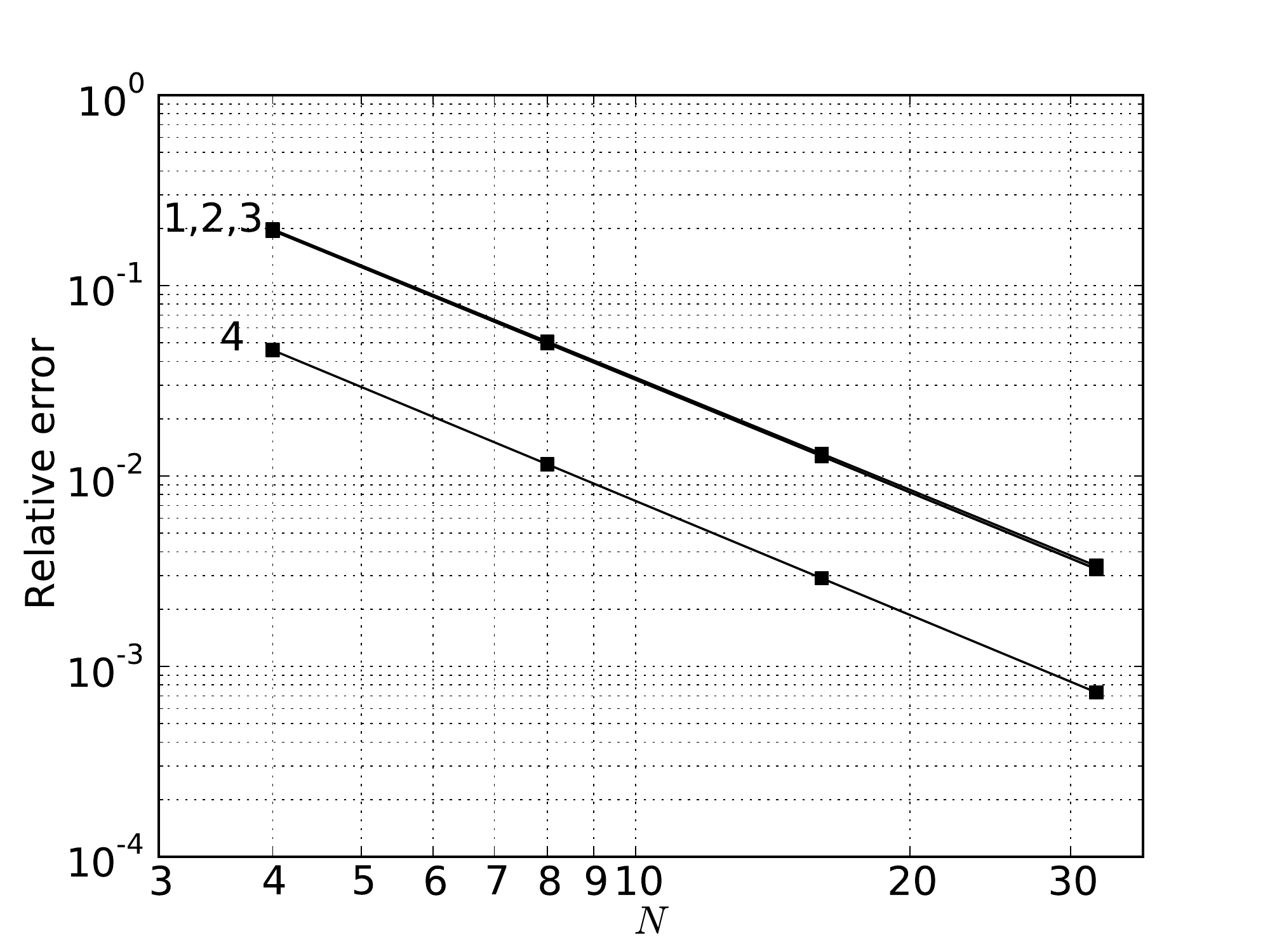}}
\subfloat[SGT\label{fig:SGT-action-error}]{\includegraphics[width=6cm]{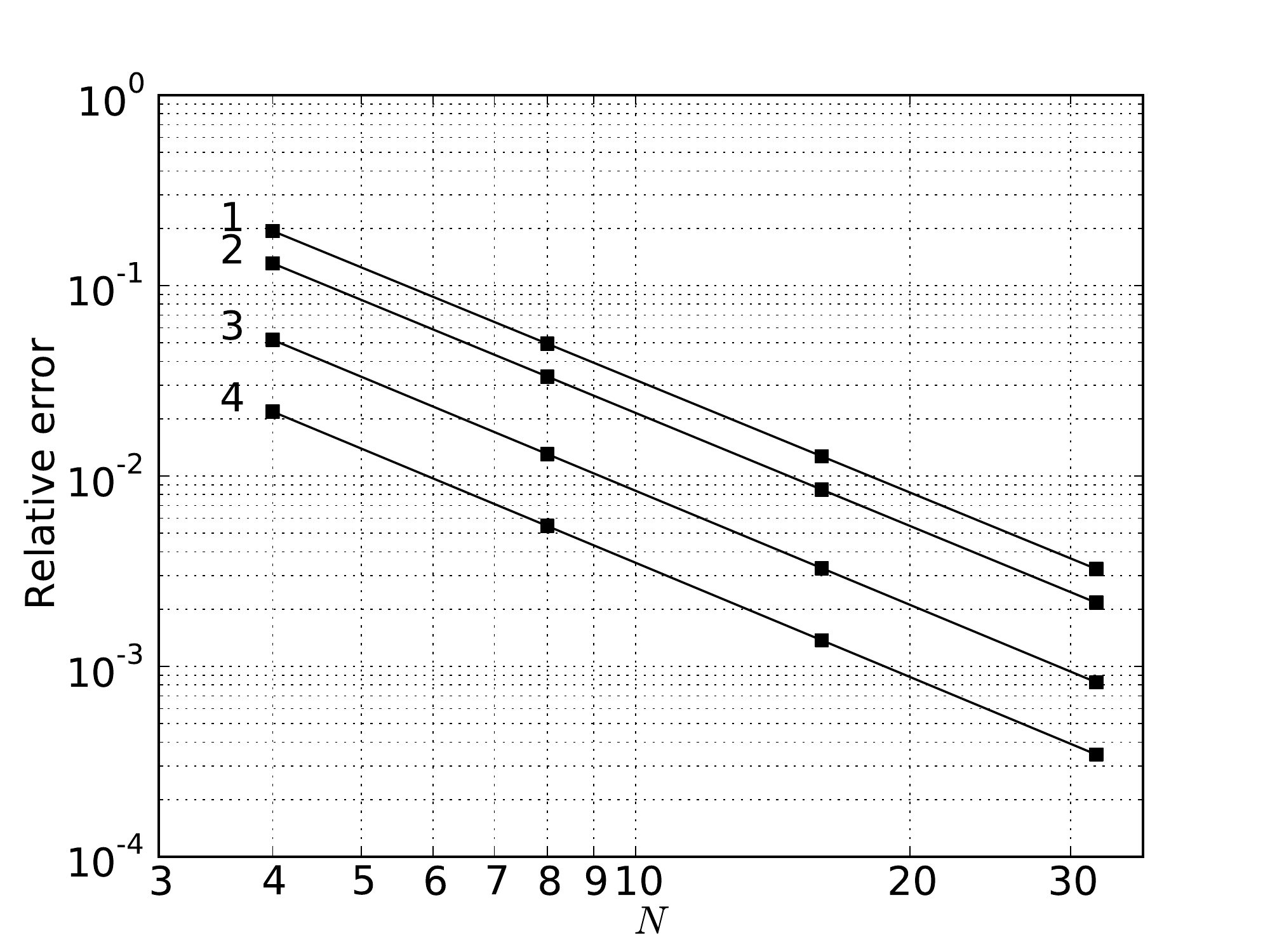}}
\caption{The relative error of the action versus the number of lattice sites
per side $N$, for the actions 1, 2, 3, 4 described in section \ref{sec:convergence}.
The squares are the simulation data points and the solid lines are
the second order polynomial fits. Errors are proportional to $h^{2}$
in all cases.\label{fig:relative-errors}}
\end{figure}

The results are displayed using double logarithmic plots in figure
\ref{fig:relative-errors} for traditional Wilson action LGT as well
as the SGT results. As expected from the estimates in \cite{HalvSor2011a},
in all cases the relative error behaves as
\[
\mbox{Relative error}\sim Ch^{2},
\]
as determined by extracting the linear coefficient of the second order
polynomial fits shown in the figures. Note that while the convergence
exponent of $h$ is the same in all cases, the prefactor $C$ is smaller
in the SGT cases involving time-independent fields, due to its finer
spatial discretization for the same $N$. Where time-dependence is
involved, the errors coincide since the time-discretization we have
chosen for this SGT simulation is of the same quality as for the LGT
simulation.

\subsection{Quantum field simulation\label{sec:MC}}

Analogous to the traditional lattice QCD simulations, we performed
parallel $SU(2)$ quantum field theory Monte Carlo simulations for
$N=8$. In this case, the edge matrices are sampled directly without
reference to a gauge field and lattice constant value. Therefore,
the physical size of the simulation domain is unknown prior to experimental
comparisons. All dimensional observable quantities are automatically
calculated in units of powers of the lattice constant $h$.

As is customary, it is a Monte Carlo simulation using the Metropolis
algorithm to generate a Markov chain of gauge field configurations
that are distributed according to the Boltzmann weight $\exp(-S)$.
Each Monte Carlo step involves randomization of some edge $SU(2)$
matrices, which is done by multiplication of a small $\mathfrak{su}(2)$
algebra matrix, together with a Metropolis step for acceptance/rejection
of the update. The algorithm adapted itself to drive the MC acceptance
rate towards $1/2$. Monte Carlo convergence tests were done and high
quality error estimates were made using data blocking \cite{FlyvPet1989}.
In addition, convergence was verified subjectively by inspection of
the time series for observable values with their accompanying distributions,
as well as time series for cumulative averages. The data blocking
error estimates were found to be smaller than the displayed data points
in all the plots.

\begin{figure}
\begin{center}
\includegraphics[width=10cm]{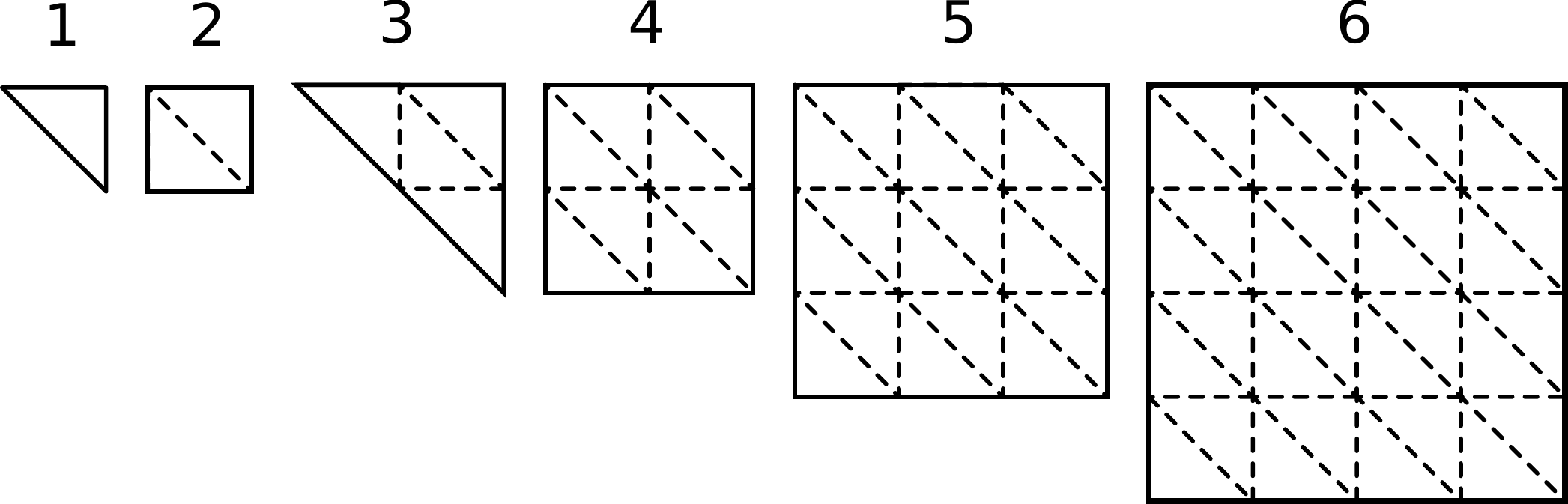}
\caption{The simulated Wilson loops shapes correspond to the outer edges of
these figures. They lie in the $xy$, $yz$ and $zx$ planes.\label{fig:Wilson-loop-shapes}}
\end{center}
\end{figure}

We simulated at different values of $\beta$, at each of which we
measured the average action density $S/N^{4}$, and a list of different
Wilson loops shown in figure \ref{fig:Wilson-loop-shapes}, all of
which are gauge-invariant quantities. For each Wilson loop shape,
we average over all possible loop positions, as well as loop orientations
in the $xy$, $yz$ and $zx$ planes. For a given closed path $\mathcal{C}$,
the corresponding Wilson loop variable for gauge group $SU(n)$ is
defined as
\begin{equation}
W_{\mathcal{C}}:=\frac{1}{2}\Re\mathrm{tr}\prod_{e\in\mathcal{C}}U_{e},\label{eq:wilson-loop}
\end{equation}
which involves an ordered product of the edge matrices $\{U_{e}\}$
along the path $\mathcal{C}$.

Expectation values for any observable quantity $\mathcal{O}$, e.g.
the action density $S/N^{4}$ or a Wilson loop $W_{\mathcal{C}}$,
is given by

\begin{equation}
\langle\mathcal{O}\rangle=\frac{1}{Z}\int\left(\prod_{e}dU_{e}\right)\mathcal{O}\exp(-S),\label{eq:observable-expectation}
\end{equation}
where the partition function $Z$ is defined by
\begin{equation}
Z:=\int\left(\prod_{e}dU_{e}\right)\exp(-S).\label{eq:partition-function}
\end{equation}
The integration measure involved in these expressions is a product
of the normalized Haar integration measure for each edge group element
in the mesh. Note that the normalized Haar measure satisfies
\begin{equation}
\int_{\mathcal{G}}dU=1.\label{eq:haar-norm}
\end{equation}

To accompany these measurements, the strong (small $\beta$) and weak
(large $\beta$) coupling asymptotic behaviour
were calculated in appendix \ref{sec:limits}, using methods described
in \cite{Creutz:1984mg}. At strong coupling, this involves various
group integrals, while at weak coupling it suffices to use a thermodynamic
analogy to determine the limiting behaviour. 

\begin{figure}
\subfloat[Action density $\langle S/N^{4}\rangle$ versus $\beta$\label{fig:Action-density}]{\includegraphics[width=6cm]{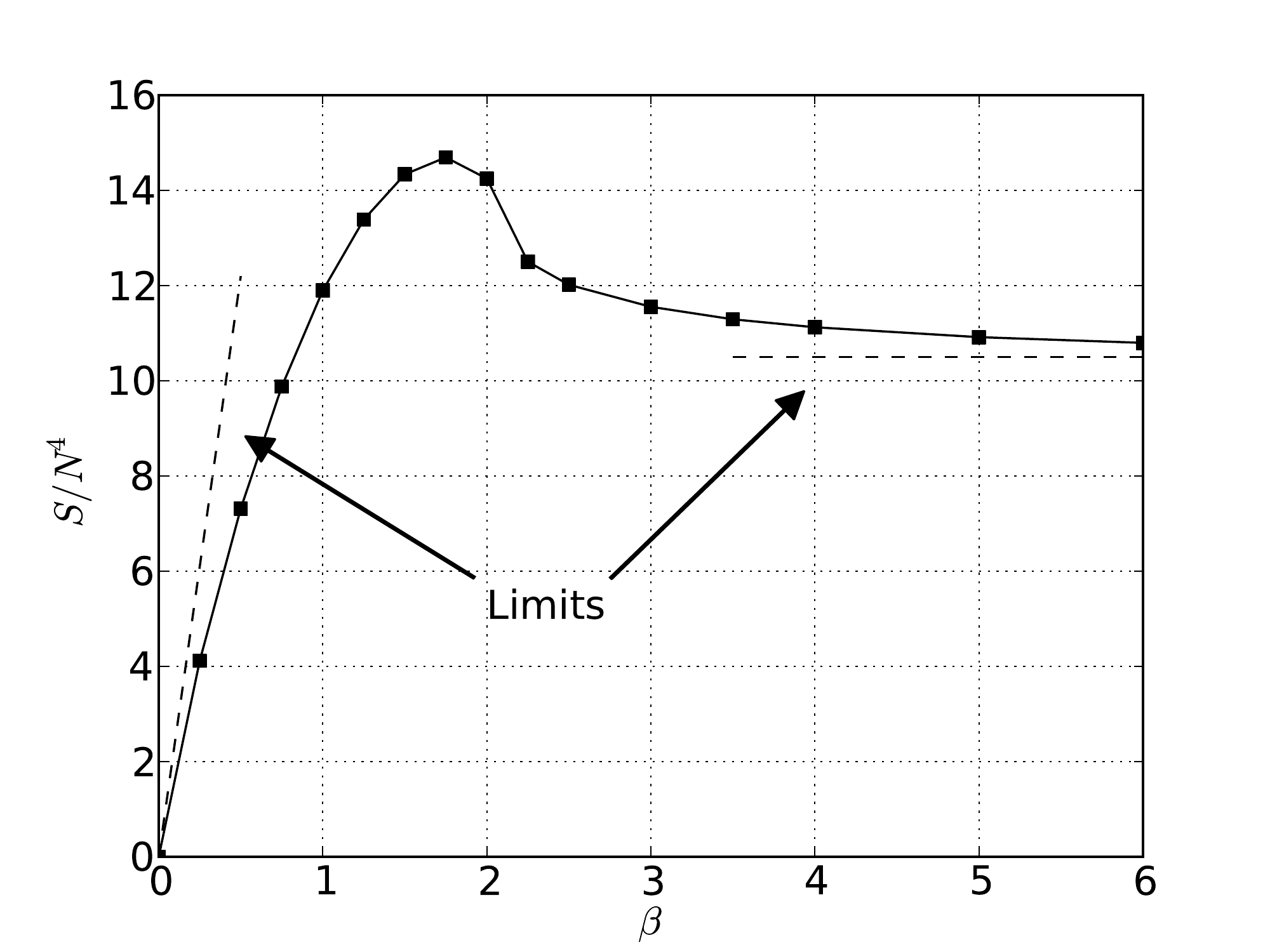}}
\subfloat[Wilson loops $\langle W\rangle$ versus $\beta$\label{fig:Wilson-loops}]{\includegraphics[width=6cm]{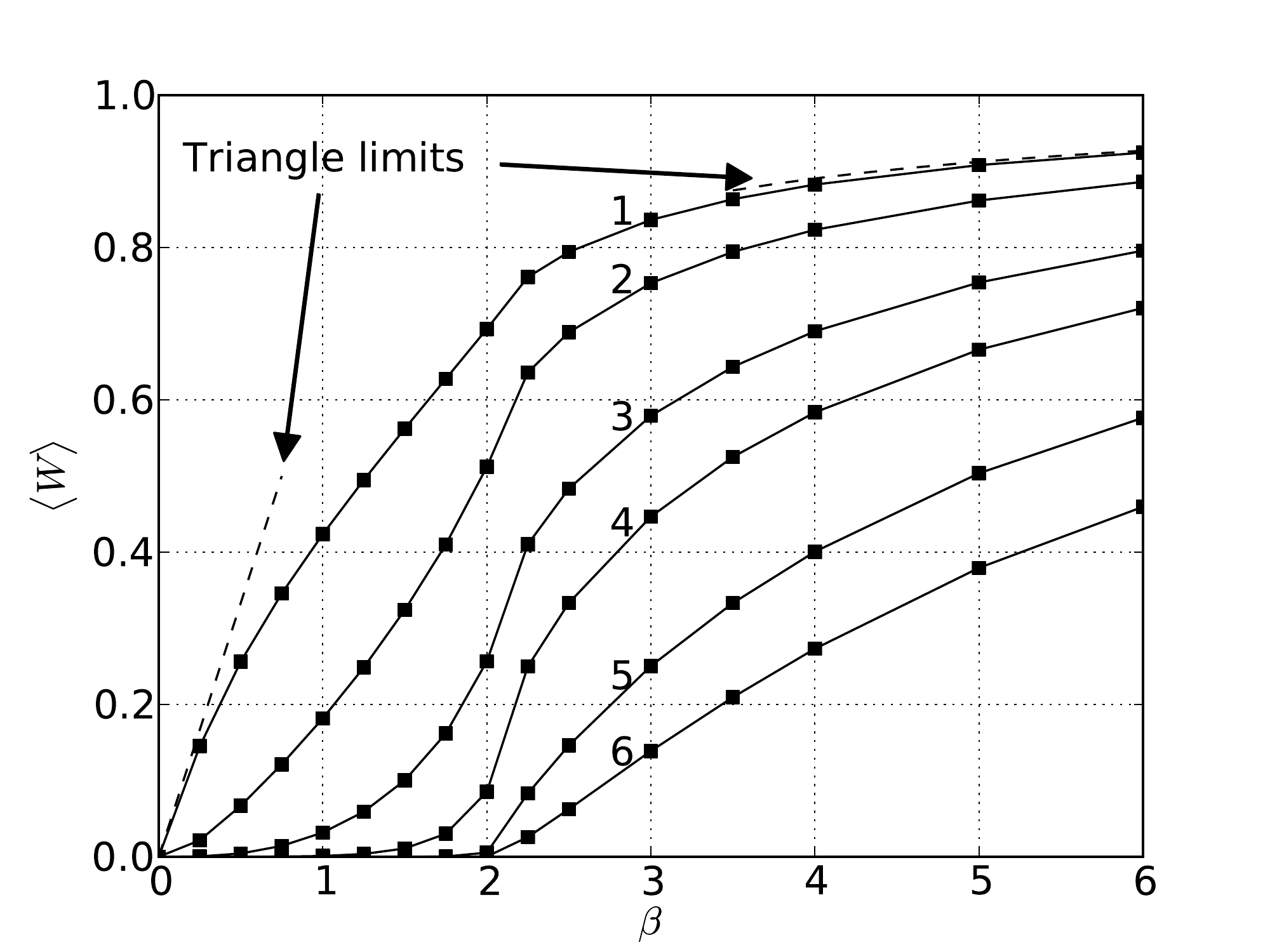}}
\caption{Plots showing the $\beta$-dependency of \subref{fig:Action-density}
the average action density $\langle S/N^{4}\rangle$ and \subref{fig:Wilson-loops}
the various Wilson loops $\langle W\rangle$ from figure \ref{fig:Wilson-loop-shapes}.
Solid squares are data points and solid lines are linear interpolations.
The strong and weak coupling asymptotes calculated in \ref{sec:limits} are included for the action
density and the elementary triangular loop. Monte Carlo errors are
smaller than the data points.\label{fig:Monte-Carlo-sim}}
\end{figure}

The simulated results for the action density and Wilson loops are
displayed in figure \ref{fig:Monte-Carlo-sim}. In figure \ref{fig:Action-density}
we can see the characteristic and nontrivial behaviour
in the medium coupling range $\beta\in(1,3)$. This coincides qualitatively
with LGT simulations \cite{Creutz:1980zw}. Only qualitative, not
exact, agreement is expected, since the physical lattice constant
will differ in each type of simulation. Compared to LGT simulations,
the behaviour at small $\beta$ deviates
more from linearity due to the nonlinear aspects of the SGT action.
In this region, the actions do not approximate the continuum action,
and differences between discrete actions are unphysical.

The Wilson loops in figure \ref{fig:Wilson-loops} show the same qualitative
behaviour as do LGT simulation results,
and approaches the calculated asymptotes nicely. Also here, the behaviour
is less linear at small $\beta$ for the same reason as stated above.
The typical strong suppression of the Wilson loops as functions of
loop area is reproduced, as expected from the area law behaviour
that indicates confinement.

\section{Conclusions\label{sec:conclusions}}

We have implemented the general SGT action on a particular simplicial
mesh, and performed Monte Carlo quantum field theory simulations that
show sensible results that are qualitatively consistent with standard
LGT simulations, as must be the case for this initial proof-of-concept
implementation.

We expect that this method will lend itself nicely to the use of mesh
refinement within quantum QCD simulations, and that this will lead
to opportunities of novel applications using nontrivial mesh structures,
e.g. in the vicinity of gluon flux tubes as mentioned in the introduction.

The nondiagonal nature of the action increases the amount of computer
work in the Metropolis step after each proposed update. However, since
the number of interactions for each elementary face is finite, the
scaling at large meshes for this model will be the same as for traditional
QCD. There might be possibilities of real-time adaptive diagonalization,
thereby increasing the algorithm efficiency throughout the initial
part of the simulation.

\appendix

\section{Simplicial complex, finite elements and mass matrices\label{sec:simp-comp}}

Consider a collection of vertexes, edges, faces, tetrahedra in 3d
space. These elementary objects are called simplexes, and the collection
of these a simplicial complex $\mathcal{T}$. For any $k$-dimensional
simplex $T_{k}$ for $1\leq k\leq3$, the boundary $\partial T_{k}$
is a union of $(k-1)$-dimensional simplexes. Consult \cite[Section 5.1]{christiansen2011}
for a precise definition. In our construction, we assume that this
spatial simplicial complex spans the spatial domain $S$. The vertexes,
edges, faces, and tetrahedra according to dimension, and are labeled
$i$, $e$, $f$, and $T$ respectively. The symbol $T$ will be used
for simplexes of any dimension.

In order to expand this to a 4d spacetime simplicial complex $\mathbb{T}$,
consider a uniform time-discretization with a time-spacing $\Delta t$.
The simplicial complex $\mathcal{T}$ is then repeated at each discrete
time step value $\tau$. For each such $\tau$, we define additional
simplexes for our $\mathbb{T}$ by extruding each simplex of $\mathcal{T}$
along the time interval $[\tau,\tau+\Delta t]$. As the basic building
block in classical 3d FEM theory is a tetrahedron $T$, the basic
building block in this extended FEM version is $T\times I_{\tau}$,
where $I_{\tau}=[\tau,\tau+\Delta t]$, i.e. a time-extrution of a
tetrahedron. Temporal edges are generated by extruding 3d vertices,
and temporal faces by extruding 3d edges.

The space of Whitney k-forms on $\mathcal{T}$ ($T$) is denoted $W^{k}(\mathcal{T})$
($W^{k}(T)$), with canonical basis $(\lambda_{T})$, $T$ ranging
over the set of $k$-dimensional simplexes in $\mathcal{T}$ \cite{whitney57}.
The 0-forms $\lambda_{i}$ are the barycentric coordinate maps for
each vertex $i$. In other words, it is the piecewise affine map taking
the value 1 at the vertex $i$ and 0 at other vertices. For an edge
$e=\{i,j\}$, with orientation $i\rightarrow j$, the associated Whitney
1-form is defined by 
\begin{equation}
\lambda_{e}:=\lambda_{ij}:=\lambda_{i}\nabla\lambda_{j}-\lambda_{j}\nabla\lambda_{i}.\label{eq:edge-basis}
\end{equation}
For a face $f=\{i,j,k\}$, whose orientation is $i\rightarrow j\rightarrow k$,
the associated Whitney 2-form is defined by 
\begin{equation}
\lambda_{f}:=\lambda_{ijk}:=2\left(\lambda_{i}\nabla\lambda_{j}\times\nabla\lambda_{k}+\lambda_{j}\nabla\lambda_{k}\times\nabla\lambda_{i}+\lambda_{k}\nabla\lambda_{i}\times\nabla\lambda_{j}\right).\label{eq:face-basis}
\end{equation}

In the 4d spacetime FEM setting, these basis $k$-forms are extended
to be piecewise affine in time and are denoted $(\Lambda_{T(\tau)})$,
i.e. 
\[
\lambda_{T}\rightarrow\Lambda_{T(\tau)}=\lambda_{T}\otimes P_{1}^{t},
\]
where $P_{1}^{t}$ denotes polynomials in the time variable of degree
at most one, and $T(\tau):=(\tau,T)$ denotes the spatial simplex
$T$ at temporal node $\tau$. More precisely, $\Lambda_{T(\tau)}$
is the piecewise affine function in time, taking the value $\lambda_{T}$
at $\tau$ and 0 at the other temporal nodes. In addition, we define
temporal basis edge and face functions.

To every vertex $i$ in the spatial mesh there are temporal edges
$e_{t}(\tau)=\{i_{\tau},i_{\tau+\Delta t}\}$, where $i_{\tau}:=i(\tau)$.
The temporal basis edge function attached to $e_{t}(\tau)$ is then
the piecewise constant function in time defined by 
\[
\Lambda_{e_{t}(\tau)}(t)=\begin{cases}
\lambda_{i}\circ\pi\frac{1}{\Delta t}dt, & t\in[\tau,\tau+\Delta t]\\
0, & \text{otherwise}.
\end{cases}
\]
where $\pi$ is the canonical projection onto the space $S$, 
\[
\pi:\mathbb{M}=\mathbb{R}\times S\rightarrow S,
\]
and $dt$ is the standard basis one-form in the temporal direction.

To every spatial edge $e$ there are corresponding temporal faces
$f_{t}(\tau)=e\times I_{\tau}$. The temporal basis face function
attached to $f_{t}(\tau)$ is then the piecewise constant function
in time defined by 
\[
\Lambda_{f_{t}(\tau)}(t)=\begin{cases}
\lambda_{e}\circ\pi\wedge\frac{1}{\Delta t}dt, & t\in[\tau,\tau+\Delta t]\\
0, & \text{otherwise}.
\end{cases}
\]

In addition to these basis functions, we must define mass matrix elements.
Let $m_{TT'}$ denote the classical 3d mass matrices for spatial Whitney
elements 
\[
m_{TT'}=\int_{S}\lambda_{T}\cdot\lambda_{T'},
\]
where $T$, $T'$ are $k$-dimensional simplexes, and ($\cdot$) denotes
the scalar product of alternating forms.

In the definition of the SGT action we use the generalization 
\[
M_{T(t)T'(\tau)}=\int_{\mathbb{M}}\Lambda_{T(t)}\cdot\Lambda_{T'(\tau)}.
\]
This generalization can be expressed through the classical mass matrices
by performing the time integration explicitly. Thus, let $T$ be a
spatial tetrahedron and $I_{\tau}=[\tau,\tau+\Delta t]$. Considering
now only this time interval, the piecewise affine function taking
the value 1 at time $\tau$ and 0 at time $\tau+\Delta t$ is given
by 
\[
p_{\tau}(t)=1-\frac{t-\tau}{\Delta t}.
\]
The analogous function for the temporal node $\tau+\Delta t$ on the
same time interval is given by 
\[
p_{\tau+\Delta t}(t)=\frac{t-\tau}{\Delta t}.
\]
Restricted to the basic building block $T\times I_{\tau}$, we therefore
get 
\[
M_{f(\tau)f'(\tau)}(T\times I_{\tau})=\int_{T\times I_{\tau}}\Lambda_{f(\tau)}\cdot\Lambda_{f'(\tau)}=\int_{I_{\tau}}p_{\tau}^{2}\int_{T}\lambda_{f}\cdot\lambda_{f'}=\frac{1}{3}\Delta tm_{ff'}(T),
\]
 
\[
\begin{split}
M_{f(\tau)f'(\tau+\Delta t)}(T\times I_{\tau})&=\int_{T\times I_{\tau}}\Lambda_{f(\tau)}\cdot\Lambda_{f'(\tau+\Delta t)}=\\
&=\int_{I_{\tau}}p_{\tau}p_{\tau+\Delta t}\int_{T}\lambda_{f}\cdot\lambda_{f'}=\frac{1}{6}\Delta tm_{ff'}(T),
\end{split}
\]
 
\[
\begin{split}
M_{f(\tau+\Delta t)f'(\tau+\Delta t)}(T\times I_{\tau})&=\int_{T\times I_{\tau}}\Lambda_{f(\tau+\Delta t)}\cdot\Lambda_{f'(\tau+\Delta t)}=\\
&=\int_{I_{\tau}}p_{\tau+\Delta t}^{2}\int_{T}\lambda_{f}\cdot\lambda_{f'}=\frac{1}{3}\Delta tm_{ff'}(T).
\end{split}
\]
Similarly, the mass matrix element corresponding to the temporal face
basis is given by 
\[
M_{f_{t}(\tau)f_{t}'(\tau)}(T\times I_{\tau})=\int_{T\times I_{\tau}}\Lambda_{f_{t}(\tau)}\cdot\Lambda_{f_{t}'(\tau)}=\frac{1}{\Delta t}\int_{T}\lambda_{e}\cdot\lambda_{e}'=\frac{1}{\Delta t}m_{ee'}(T).
\]

\section{Strong and weak coupling limits\label{sec:limits}}

\subsection{Strong coupling limit\label{sub:strong-coupling}}

Here we will show some details regarding the calculation of the strong
coupling limits of the elementary triangular Wilson loop. We will
use the following integrals over $SU(2)$ group space \cite{Creutz:1984mg}
\begin{equation}
\begin{split}
\int dUU^{\alpha\beta}=0,\quad\int dUU^{\alpha_{1}\beta_{1}}U^{\dagger\beta_{2}\alpha_{2}}=\frac{1}{2}\delta^{\alpha_{1}\alpha_{2}}\delta^{\beta_{1}\beta_{2}},\\
\int dUU^{\alpha_{1}\beta_{1}}U^{\alpha_{2}\beta_{2}}=\frac{1}{2}\epsilon^{\alpha_{1}\alpha_{2}}\epsilon^{\beta_{1}\beta_{2}},\label{eq:group_int}
\end{split}
\end{equation}
where the Greek symbols are matrix indices.

In this calculation, the Wilson loop encircles an elementary spatial
triangular plaquette $P_{t}$ at time $t$. We denote this Wilson
loop by $W_{P_{t}}$. By equation \eqref{eq:wilson-loop}, it is given
by

\[
W_{P_{t}}:=\frac{1}{2}\Re\mathrm{tr}\left(U_{a}U_{b}U_{c}\right),
\]
where the plaquette $P_{t}$ is encircled cyclically by the $SU(2)$
edge matrices $U_{a}$,$U_{b}$ and $U_{c}$. Due to our choice of
distinguished points and plaquette orientations, the spatial SGT action
is given by
\[
S=\frac{\beta}{2}\sum_{f,f'}M_{ff'}\mathrm{tr}\left(U_{f}U_{f'}^{H}-U_{f}-U_{f'}^{H}+\mathbbm1\right),
\]
where the sum extends over all spatial faces at all times. Since we
are interested in small $\beta$, consider a first order truncated
Taylor expansion of the exponential in equation \eqref{eq:observable-expectation},
i.e. 
\[
\langle W_{P_{t}}\rangle\approx\frac{-\beta}{4Z_{\beta}}\int\left(\prod_{e}dU_{e}\right)\Re\mbox{tr}\left(U_{a}U_{b}U_{c}\right)\sum_{f,f'}M_{ff'}\mbox{tr}\left(U_{f}U_{f'}^{H}-U_{f}-U_{f'}^{H}+\mathbbm1\right).
\]
By the properties of the $SU(2)$ integration measure, terms involving
integration over odd powers of link matrices vanish. Therefore, nonvanishing
contributions to the integral only come from terms where either $f$
and/or $f'$ coincide with the plaquette $P_{t}$. The $U_{f}U_{f'}^{H}$
doesn't contribute. Indeed, if either $f$ or $f'$ differ from $P_{t}$,
we such a term includes an integral over a single power, which vanishes.
If on the other hand $f=f'=P_{t}$, we have $U_{f}U_{f}^{H}=\mathbbm1$
which again leads to an integral over a single power and thus vanishes.
This is also the case for the constant term in the parenthesis.

We are left with
\[
\langle W_{P_{t}}\rangle\approx\frac{\beta}{4Z_{\beta}}\Re\int\left(\prod_{e}dU_{e}\right)\mbox{tr}\left(U_{a}U_{b}U_{c}\right)\sum_{f,f'}M_{ff'}\mbox{tr}\left(U_{f}+U_{f'}^{H}\right),
\]
where we have moved the real part operator $\Re$ outside of the integral.
Contributions only come when at least one of $f,f'$ coincide with
$P_{t}$. Therefore, by the properties of the particular mesh we have
constructed,
\begin{equation*}
\begin{split}
\langle W_{P_{t}}\rangle\approx\frac{\beta}{4Z_{\beta}}&\left(M_{P_{t}P_{t}} + M_{P_{t}P_{t+1}} + M_{P_{t}P_{t-1}}\right)\times\\
&\times\Re\int\left(\prod_{e}dU_{e}\right)\mbox{tr}\left(U_{a}U_{b}U_{c}\right)\mbox{tr}\left(U_{P_{t}} + U_{P_{t}}^{H}\right).
\end{split}
\end{equation*}
Using $U_{P_{t}}:=U_{a}U_{b}U_{c}$ and the $SU(2)$ integration formulas
\eqref{eq:group_int}, we get
\[
\langle W_{P_{t}}\rangle\approx\frac{\beta}{2}\left(M_{P_{t}P_{t}}+M_{P_{t}P_{t+1}}+M_{P_{t}P_{t-1}}\right)=\frac{2}{3}\beta,
\]
where we have used $Z\approx1$ for small $\beta$. The last equality
follows from the particular mass matrix element values produced by
our choice of simplicial lattice.

A similar calculation, only slightly more involved because several
faces are involved, can be performed to determine the strong coupling
limit of the action. Approximations of higher order in $\beta$ can
be found by including higher order terms in the Taylor expansion of
the exponential.

\subsection{Weak coupling\label{sub:weak-coupling}}

In order to determine the weak coupling limit of the action density,
we simple follow a thermodynamic analogy described in \cite{Creutz:1984mg}.
At large $\beta$, the system is described well by a gaussian partition
function approximation. This corresponds to a free theory, and we
can find the weak coupling limit of the action by distributing an
amount $kT/2=1/2\beta$ of energy among all the degrees of freedom
in the theory. We have seven edges for each building block cube, each
of which contributes three degrees of freedom (the number of generators
of $SU(2)$). To obtain the action, we multiply by $\beta$, which
results in
\begin{equation}
S_{SGT}\rightarrow\beta\times\frac{1}{2\beta}\times7\times3=\frac{21}{2}N^{4},\quad\mbox{as }\beta\rightarrow\infty.\label{eq:action-dens-weak}
\end{equation}
This result can be used to determine the same limit of the triangular
Wilson loop in the $\alpha\beta$-plane. We have
\[
\langle W_{1}\rangle=1-\frac{a^{4}}{16}\langle\mbox{tr}(F_{\alpha\beta}^{2}\rangle,
\]
where there is no sum over the spacetime indices. The antisymmetric
field strength has six independent spacetime components. By the equipartitioning
of the euclidean energy among these degrees of freedom, we have
\[
\langle\mbox{tr}(F_{\alpha\beta}^{2}\rangle=\frac{1}{6}\langle\mbox{tr}(F_{\mu\nu}F^{\mu\nu})\rangle=\frac{2g^{2}}{6}\langle\frac{S_{\mbox{SGT}}}{N^{4}}\rangle=\frac{42g^{2}}{12}.
\]
Now using $\beta=2/g^{2}$, we get

\begin{equation}
\langle W_{1}\rangle=1-\frac{21}{48\beta}.\label{eq:triangle-weak}
\end{equation}

\section{Computer implementation\label{sec:program}}

Our computer implementation of the simplicial lattice and accompanying
SGT action consists of object-oriented C++ code, using MPICH2 \cite{mpich2}
for parallelization, running on a quadruple CPU run-of-the-mill modern
workstation computer. The data structures involved are reminiscent
of what is used in implementations of the finite element method. This
involves different types of mass matrix and connectivity information
for elements of the simplicial mesh. The parallelization consisted
of running independent simulations on each node, and averaging the
results. We used the yarn2 algorithm from the TINA pseudo-random number
generator \cite{tina}, which is designed for use in parallelized
algorithms.  Although the edge matrix randomization appeared to perform
stably enough for our purposes, we regularly did projections of the
edge matrices onto $SU(2)$ as a precautionary measure.

\bibliographystyle{elsarticle-num}
\bibliography{ref}

\end{document}